\documentclass[aps,prl,reprint,floatfix]{revtex4-2}

\usepackage{hyperref,graphicx,bm,amsmath,color,comment} 
\usepackage{braket}
\usepackage{float}
\usepackage[dvipsnames]{xcolor}

\begin{document}

\title{Crucial role of Fe in determining the hard magnetic properties of Nd$_2$Fe$_{14}$B}
\author{Juba Bouaziz}
\email{j.bouaziz@fz-juelich.de}
\affiliation{Department of Physics, University of Warwick,
Coventry CV4 7AL, UK}
\affiliation{Peter Gr\"{u}nberg Institut and Institute for Advanced 
Simulation, Forschungszentrum J\"{u}lich \& JARA, D-52425 J\"{u}lich, Germany}
\author{Christopher E.\ Patrick}
\affiliation{Department of Materials, University of Oxford,
Parks Road, Oxford OX1 3PH, UK}
\author{Julie B. Staunton}
\email{j.b.staunton@warwick.ac.uk}
\affiliation{Department of Physics, University of Warwick,
Coventry CV4 7AL, UK}

\date{\today}

\begin{abstract}
Nd$_2$Fe$_{14}$B's unsurpassed, hard magnetic properties for a wide
range of temperatures result from a combination of a large volume magnetization 
from Fe and a strong single-ion anisotropy from Nd. Here, using finite temperature 
first-principles calculations, we focus on the other crucial roles played by the Fe 
atoms in maintaining the magnetic order on the Nd sublattices, and hence the large 
magnetic anisotropy, and directly generating significant uniaxial anisotropy at high 
temperatures. We identify effective spins for atomistic modelling from the material's 
interacting electrons and {quantify pairwise and higher order, non-pairwise magnetic interactions among them. We find the Nd spins couple most strongly to spins on sites belonging to two specific Fe sublattices, 8$j_1$, 8$j_2$. Moreover the Fe 8$j_1$ sublattice also provides the electronic origin of the unusual, nonmonotonic temperature dependence of the anisotropy 
of Y$_2$Fe$_{14}$B.} Our work provides 
atomic-level resolution of the properties of this fascinating magnetic material.
\end{abstract}

\maketitle 

The elemental lanthanides show remarkable magnetic properties 
deriving from their partially-filled shells of atomic-like 4$f$ electrons.
%
%
However, direct exploitation of these properties is hindered
by low magnetic ordering temperatures.
No elemental lanthanide retains its magnetism at room temperature,
with the highest Curie temperature $T_c$ being 292~K for Gd~\cite{Elliottbook}.
Combining the lanthanides with other elements 
can strengthen the magnetic interactions and allow ordering
to persist to higher temperatures.
The most successful example of this paradigm is the rare-earth/transition-metal
(RE-TM) family of permanent magnets~\cite{Coey2011}.
Specifically, Nd-Fe-B demonstrates exceptional magnetic strength over a wide range
of temperatures.
Having revolutionized computer hard disk technology in the last century,
Nd-Fe-B is again under intense investigation owing to its use in 
electric vehicle motors and renewable energy turbines~\cite{Hirosawa2018}.
%
%
%
%

The RE-TM magnetic interactions are most simply described in terms
of the exchange field $\boldsymbol{B}_\mathrm{exch}$.
In this picture, the TM-3$d$ electrons produce an effective magnetic
field which couples to the spin magnetic moments of the RE ions.
A minimal model to describe the RE ions and the high magnetic anisotropy they generate combines this exchange field
with the interaction with an external field
$\boldsymbol{B}_{\mathrm{ext}}$
and the crystal
field $\hat{V}_{\mathrm{CF}}$, which describes 
the (predominantly) electrostatic interaction of
the 4$f$ charge cloud with its environment~\cite{richter1998band,Kuzmin2007ChapterTT,patrick2019first,Patrick2019}:
\begin{equation}
\mathcal{H}_\mathrm{RE} =  2\mu_B\,\hat{\boldsymbol{S}}\cdot\boldsymbol{B}_\mathrm{exch}
+ \mu_B\,(\hat{\boldsymbol{L}} + 2\hat{\boldsymbol{S}})\cdot\boldsymbol{B}_{\mathrm{ext}} 
+ \hat{V}_{\mathrm{CF}}.
\label{eq.RE}
\end{equation}
$\hat{\boldsymbol{S}}$ and 
$\hat{\boldsymbol{L}}$
are the total spin and orbital angular momentum operators.
Values of the exchange field can be extracted by fitting Eq.~\ref{eq.RE}
to experimental data obtained in inelastic neutron scattering (INS)
or magnetization measurements.
Experimental estimates of
$\boldsymbol{B}_\mathrm{exch}$ are far stronger
than fields achievable in the laboratory 
($\mu_BB_{\mathrm{exch}}/k_B \gtrsim 
300~K$~\cite{Herbst1991}, i.e.\
$B_{\mathrm{exch}} \gtrsim$ 450~T)
as required to
maintain magnetic order above room temperature.

Going beyond a phenomenological understanding of RE ordering
requires an atomistic picture of the magnetic interactions among effective spins.
Nd$_2$Fe$_{14}$B has a tetragonal crystal structure with 68 atoms per unit cell (\cite{Herbst1991}~\footnote{See Supplemental Material for a picture of the crystal structure, site-resolved local, electronic spin-polarized densities of states, further information about multi-spin interactions, Y$_2$Fe$_{14}$B magnetic properties, the $4f$-atomic Hamiltonian and numerical values of magnetic interactions between pairs of sites (see, also, references~\cite{Isnard1995,huang1987first,bolzoni19873,stevens1952,enkovaara2010,Bauer:2014,Vosko:1980} therein).}). 
The RE atoms occupy two crystallographically distinct sites (RE$_{4f}$ and RE$_{4g}$), 
which (together with Fe$_{4c}$ and B$_{4g}$ atoms) 
form planes encapsulating the remaining 5 Fe sublattices (4$e$, 8$j_1$, 8$j_2$, 16$k_1$, 16$k_2$).
For the Nd sites the spins come from the localized f-electrons but for the TM sites the local effective spins, or local moments, emerge from the material's itinerant electron fluid~\cite{Gyorffy_1985}. Spin-polarized regions at atomic sites form from co-operative behavior of the valence electrons and at finite temperatures
 their orientations fluctuate on relatively long time scales compared to the remaining electronic degrees of freedom.
 These local magnetic moments are the pertinent, effective spins for the TM aspect of the atomistic modelling.

\begin{figure*}
\includegraphics[width=1.0\textwidth]{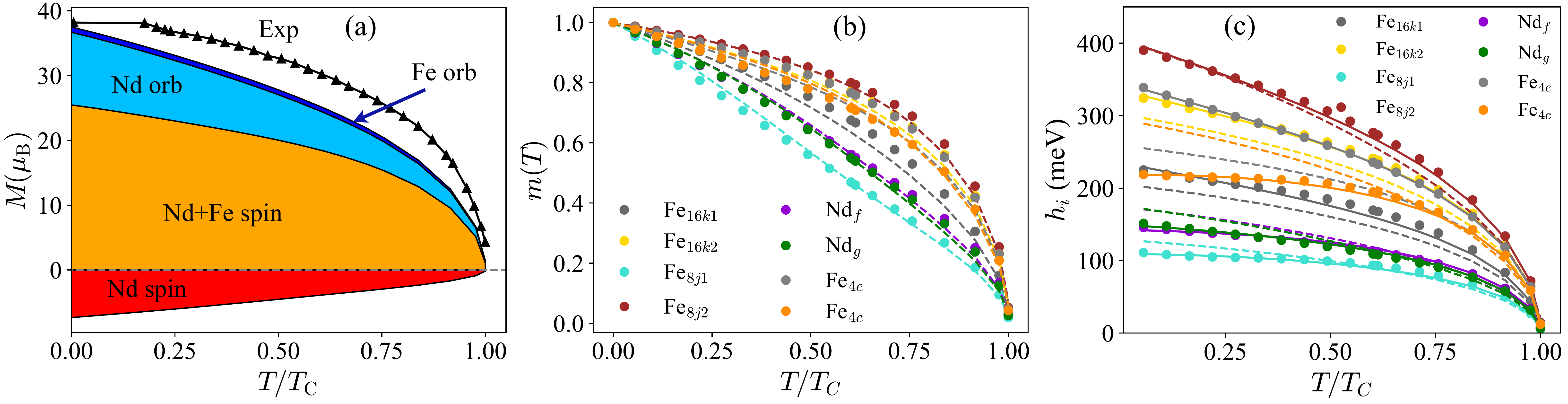}
\caption{(a) Nd$_2$Fe$_{14}$B's magnetization versus $(T/T_{c})$  
from DLM-DFT calculations compared to experiment~\cite{hirosawa1986magnetization}. (b) Sub-lattice resolved magnetic order parameters and (c) Weiss fields. 
The dots indicate the full DLM-DFT results, dashed lines from a pair-wise interaction model and continuous lines from a fit of the DLM-DFT results to the model discussed in the text 
Eq.~\eqref{eq.spinmodel}.}
\label{fig.stuffvT}
\end{figure*}

A conceptually simple model assumes interactions only
between pairs of spins $(ij)$ according to the classical Heisenberg model,
$-\mathcal{J}_{ij}\hat{\boldsymbol{S_i}}\cdot\hat{\boldsymbol{S_j}}$ where
$\hat{\boldsymbol{S_i}}$ represents an effective spin. 
Previous works~\cite{liechtenstein1984exchange,liechtenstein1987local}
calculate such $\mathcal{J}_{ij}$ parameters from first principles within density-functional theory (DFT)~\cite{liechtenstein1984exchange,liechtenstein1987local}, and use them directly in atomistic spin dynamics simulations.
With the TM magnetocrystalline anisotropy (MCA) modelled as a sum of single ion-like terms, assumed to be substantial, and RE crystal field coefficients taken from experiment, the simulations can reproduce the magnetization behavior
of Nd$_2$Fe$_{14}$B, including the spin reorientation transition
at low temperature, and represent the current state-of-the-art
in modelling these magnets~\cite{Toga2016}.
Although such a pair-wise Heisenberg model is computationally straightforward to implement, it is nonetheless a clear presumption for a magnetic metal like Nd-Fe-B.
Despite the huge technical importance of the material, the role of 
``beyond Heisenberg'' itinerant electron spin features has yet to be elucidated for Nd-Fe-B. Moreover the MCA from the spin-orbit coupling of the itinerant d-electrons is also not guaranteed to be single-ion like~\cite{Staunton2006}. In this letter we quantify the significance of both these aspects and propose ways to improve atomistic spin modelling. 

The disordered local moment (DLM) picture implemented within DFT provides an appropriate {\it ab initio}
framework~\cite{Gyorffy_1985,Staunton2006,Patrick_2019}.
The approach combines statistical mechanics of the effective spins (local moments, {$\{ \boldsymbol{e_i} \}$}) and DFT, to describe the complementary evolution of 
electronic and magnetic structure as a material's temperature
is varied. Strongly correlated 4f-electron effects are treated with a parameter free, self interaction correction (SIC) approach~\cite{Perdew1981,Luders2005} which incorporates Hund's rules naturally~\cite{Patrick2018}. As such DLM-DFT can describe temperature-dependent magnetic properties of permanent magnets as shown recently for the RECo$_5$ family~\cite{Patrick2019}. The crucial RE contribution to the anisotropy is accounted by crystal field theory, calculating the CF coefficients within DFT using a robust numerical method~\cite{Patrick_2019} so that the modelling is independent of any prior fit of phenomenological parameters.   

Here we investigate the nature of magnetic order in Nd$_2$Fe$_{14}$B, sublattice-resolved, and describe the magnetic
interactions among the effective spins associated with both RE and TM sites.
We show that the interactions among the TM spins 
are influenced by the global magnetic order and its impact and link with the spin-polarised electrons of the system. This is in essence a multi-spin coupling effect.
We find significant diversity in the behavior of the Fe local moments depending on their location in the unit cell. While most TM spins are ferromagnetically-coupled, some interact antiferromagnetically with each other. This leads to some frustration and a peculiar strong suppression of magnetic order on the ${8j_{1}}$ sites {which are located roughly midway between the Nd-containing layers. We also find that the Nd spins couple most strongly to spins on sites belonging to this Fe sublattice along with those on another (8$j_2$).}
Furthermore we discover a link between this 8$j_1$ sublattice and 
the unusual non-monotonic temperature dependence
of the non-RE MCA of the isostructural
material Y$_2$Fe$_{14}$B, resolving a longstanding
a puzzle~\cite{hirosawa1986magnetization,cadogan1992analysis}.
Finally our calculation of
the anisotropy field of Nd$_2$Fe$_{14}$B across a range of temperatures agrees well with experiment and
confirms the vital role played by the Fe spins for the functionality of this champion magnet.

Apart from the local moments themselves, the central quantities
in DLM-DFT theory are Weiss fields $\{\boldsymbol{h_i}\}$ which,
analogously to the exchange field of Eq.\ref{eq.RE}, drive the ordering of the local moments.
However, unlike $\boldsymbol{B}_\mathrm{exch}$, the Weiss fields
are not phenomenological, but instead are rigorously defined by {thermal averages over the local moment orientational configurations $\{ \boldsymbol{e_i}\}$ of the magnetic energy $\Omega \{ \boldsymbol{e_i}\}$~\cite{Gyorffy_1985,Staunton2006}, i.e.
\begin{equation}
\boldsymbol{h_i} =\int\frac{3}{4\pi}\langle \Omega \rangle _{\boldsymbol
{e_i};T}\,d\boldsymbol{e_i}\;.\label{eq.weiss}
\end{equation}
where $\langle X \rangle _{\boldsymbol{e_i}}$ denotes the average of $X$ with the restriction
that the orientation of the moment on site $i$ is fixed as $ {\boldsymbol{e_i}}$ and the order parameters of the local moments,
$\{\boldsymbol{m_i} \}$ are the averages $\{\langle \boldsymbol {e_i} \rangle \}$~\cite{Gyorffy_1985,Staunton2006}.
$\langle\Omega\rangle_{\widehat{e}_{i};T}$}
is calculated from DFT~\cite{Gyorffy_1985}. Crucially no prior prescription is assumed for the form of the magnetic interactions inherent in the first-principles $\Omega$.
For a pairwise Heisenberg model, {the magnetic energy
$\Omega\{ \boldsymbol{e_i}\}= -1/2 \sum_{ij}  \mathcal{J}_{ij} \boldsymbol{e_i}\cdot
\boldsymbol{e_j}$} with Weiss fields linear in the $\{\boldsymbol{m_i}\}$,
$\boldsymbol{h_i} = \sum_{j} \mathcal{J}_{ij} \boldsymbol{m_j}$.
Consequently beyond-pairwise terms are clearly identified in DLM-DFT theory from the non-linear dependence of the Weiss fields on the $\{\boldsymbol{m_i}\}$
~\cite{Staunton:2014,Mendive2019,Boldrin:2018,Mendive2017,Mendive2020}. 
The $\{\boldsymbol{m_i}\}$ order parameters, describing an equilibrium state at a temperature $T$, are given by the self-consistent solution of Eq.\ref{eq.weiss} and $m_i= (-1/\beta h_i + \coth \beta h_i)$, the Langevin function, ($\beta= 1/k_BT$).
\begin{figure}
\hspace{-4mm}
\includegraphics[width=0.48\textwidth]{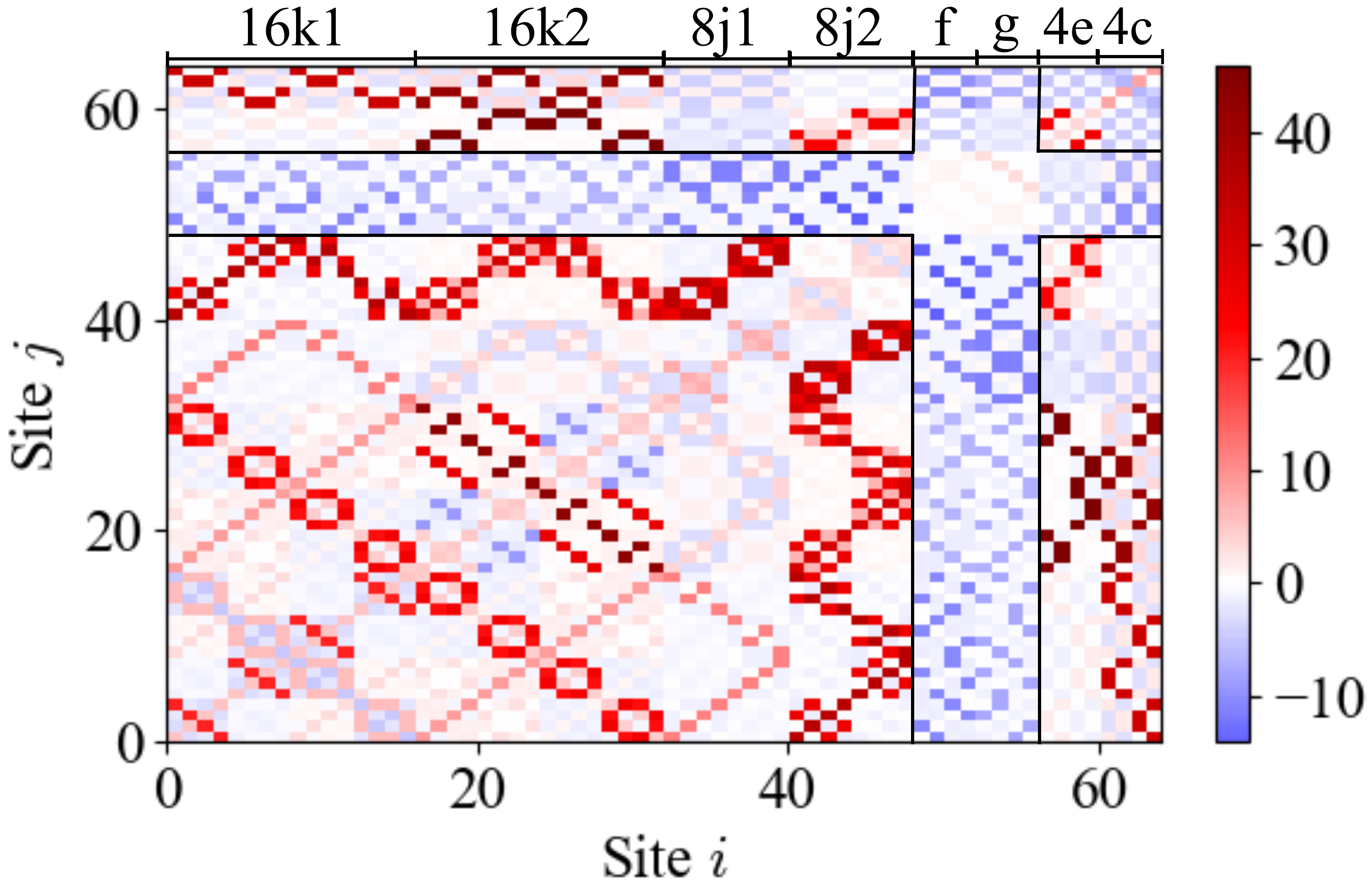}
\caption{The relative strengths of interactions between sites in the unit cell (boron sites not included), $\mathcal{J}_{ij}$, (Eq.~\ref{eq.spinmodel}) highlighting the RE-TM ones (sites 48--51 and 52--55 correspond to $4f$ and $4g$ respectively). Numerical values in meV are given in \cite{Note1} along with specific site coordinates. Red/blue color indicates FM/AF interactions.}
\label{fig.Jij}
\end{figure}

Figure~\ref{fig.stuffvT}(a) shows the magnetization  as a function of $T$ compared to experiment and resolved into the RE and TM spin and orbital components. {The magnetization is directed along $\theta=45^{\circ}$ in the $(xz)$-plane.} Full calculational details are given in the Supplemental Material~\cite{Note1} and
references~\cite{Matsumoto2014,Patrick2017,Patrick_2019,Patrick2022}.
The contribution from a particular site $i$ is found by multiplying its local moment magnitude, $\mu_i$, by the order parameter $\boldsymbol {m}_i(T)$. 
The Fe and Nd spin moments interact antiferromagnetically (AF) and order in an anti-parallel alignment in a ferrimagnetic state, but the large orbital moment of Nd, 
pointing opposite to its spin, leads to overall ferromagnetic (FM) order.
The Fe orbital moments are small ($\sim 0.05\mu_B$/atom).
The calculated $T_c$ is $1058$~K, which, although an overestimate of $473$~K in comparison to the experiment~\cite{hirosawa1986magnetization}, is reasonable for a first-principles theory which uses a mean field approximation for the statistical mechanics of the effective spins~\cite{Patrick2022}.  

On each of the six Fe and two Nd sublattices (\cite{Herbst1991,Note1} the magnetic order varies from complete, $\{m_i=1\}$, at $T=0$K to zero above $T_c$, $\{m_i=0\}$.
Figure~\ref{fig.stuffvT}(b) shows how the temperature dependence of  magnetic order varies across the sublattices. 
The Nd sublattices disorder more quickly
than all the Fe sublattices except the $8j_1$ one. Complementary information in Fig.~\ref{fig.stuffvT}(c) shows that Weiss fields, $\{\boldsymbol{h_i}\}$, promote strong ordering when large and have considerable sublattice variation, notably the factor $\sim$4 difference between the $8j_1$ and $8j_2$ sites.
Analysis of $\{\boldsymbol{h_i}\}$, Eq.\ref{eq.weiss}, reveals the presence and importance of interactions that fall outside those of a Heisenberg-like model. 
For such a pairwise model the $\mathcal{J}_{ij}$ interactions (Fig.~\ref{fig.Jij}), directly obtained from the Weiss fields for small values of the $\{m_i\}$,  are used to construct the model's Weiss fields and $\{m_i\}$ at all $T$  (dashed lines in Fig.~\ref{fig.stuffvT}(c)). There are large discrepancies from the full {\it ab initio} DLM-DFT data away from $T_c$, leading us to propose a more realistic representation of the interactions which is straightforward to incorporate into atomistic spin modelling of the magnet's properties. It leads to a magnetic energy per unit cell
\begin{equation}
\bar{\Omega} = -\frac{1}{2}\sum_{ij}\mathcal{J}_{ij}\boldsymbol{m}_{i}\cdot \boldsymbol{m}_{j}-\frac{1}{4}\sum_{i} \mathcal{B}_{I} (\boldsymbol{m}_{i} \cdot \boldsymbol{M})^2,
\label{eq.spinmodel}
\end{equation}
where $i$, $j$ run over the sites in the unit cell, $I$ denotes one of the 8 sub-lattices to which the site $i$ belongs and $\boldsymbol{M}$ is the total magnetization per unit cell, $\boldsymbol{M} = \sum_{i}\mu_{i}\boldsymbol{m}_{i}$ where the order parameters on the RE sites are anti-parallel to the TM sites for the ferrimagnetic state. The second, higher order term captures the effect of the overall spin-polarization of the electronic structure on the effective interactions between the local moments. Computing Weiss fields from this expression fits the DLM-DFT calculations very well as shown by the full curves in Fig.~\ref{fig.stuffvT}(c) and $\bar{\Omega}$ closely approximates $\langle\Omega\rangle_T$.
Table~\ref{ho_param_dlm} lists the $\mathcal{B}_I$ parameters that measure the sublattice-dependent size of these higher order, multi-spin terms.

\begin{table}[H]
	\begin{center}
		\begin{tabular}{lcccccccccc}
			\hline 
            System             & $4c$   & $4e$ & $8j1$ & $8j2$ & $16k_1$ & $16k_2$ & $R_{f}$ & $R_{g}$ \\
			\hline  
			Nd$_{2}$Fe$_{14}$B & -15.42 & 14.31 & -5.06 & -1.38 & 3.82 & 4.44 &  -2.53  &  -1.41  \\
			\hline
            Y$_{2}$Fe$_{14}$B  & -13.68 & 9.91 & -4.07 &  1.27 & 4.25 & 4.07 &  0.0  &  0.0  \\
			\hline
		\end{tabular}
		\caption{Effective, multi-spin interaction constants (in $\mu$eV), $\mathcal{B}_{I}$, for Nd$_{2}$Fe$_{14}$B and Y$_{2}$Fe$_{14}$B.}
        \label{ho_param_dlm}
	\end{center}
\end{table} 
Fig.~\ref{fig.Jij} shows the relative strengths of the $\mathcal{J}_{ij}$ interactions between pairs of sites. They are represented on a $64\times64$ grid (56 Fe sites and 8 RE sites and arranged according to sublattice). Numerical values are given as Supplemental Material~\cite{Note1}. Assuming a range less than roughly 5\AA, they can be directly used in atomistic spin simulations together with the terms from Table~\ref{ho_param_dlm}. The figure illustrates the vital importance on the RE magnetic ordering of the hexagonal nets of Fe atoms~\cite{Herbst1991,Note1} from the $k_1$, $k_2$ {and notably sites on the} $j_1$ and $j_2$ sublattices. Indeed the largest contributions to the Weiss fields at the RE sites originate from the {$j_1$ and $j_2$} sublattices.

 The TM-TM interactions are particularly varied ranging from FM (red) for the majority to AF (blue). The $j_1$ sites have AF interactions with $e$, $c$ and RE sites and strong FM ones with $j_2$ sites. This frustration  drives this sublattice's aversion to magnetic order.  The diversity of the interactions stems from the profound effect that atom coordination and spacing of Fe atoms in a metallic material has on its magnetism.  The archetype for this quintessentially itinerant electron effect is fcc Fe where squeezing the lattice turns ferromagnetic order anti-ferromagnetic and then destroys it~\cite{Kubler81,Pinski86}. 
 
The diverse nature of the magnetic order on the six Fe sub-lattices also has an impact on the intrinsic MCA generated by the system's itinerant valence electrons. 
As found for other TM metal magnets~\cite{Staunton2006,Matsumoto2014}, a simple fit in terms of a single ion model is unsatisfactory {and, as found for other itinerant electron magnets, two-ion type terms should be included in the model~\cite{Staunton2006,cuadrado2021first,evans2020temperature}}. 
Furthermore, on general grounds, modelling the MCA as a sum of single ion anisotropy terms must be done extremely carefully.
The various Fe sites have different crystallographic point symmetries, and
their unique symmetry axes do not necessarily point along the $c$ direction{~\cite{miura2014magnetocrystalline}}. There are important implications for atomistic spin dynamics simulations{~\cite{Toga2016,Gong2019,cuadrado2021first}}
where it is not correct to assign a single ion anisotropy to each Fe atom with the same angular dependence and same symmetry axes.
Rather a simpler and more rigorous alternative would be to compute an anisotropy energy based on the vector sum of 
all moments in the same sublattice so that a single symmetry axis is recovered~\cite{Kuzmin2007ChapterTT}.

\begin{figure}[H]
\includegraphics[width=0.48\textwidth]{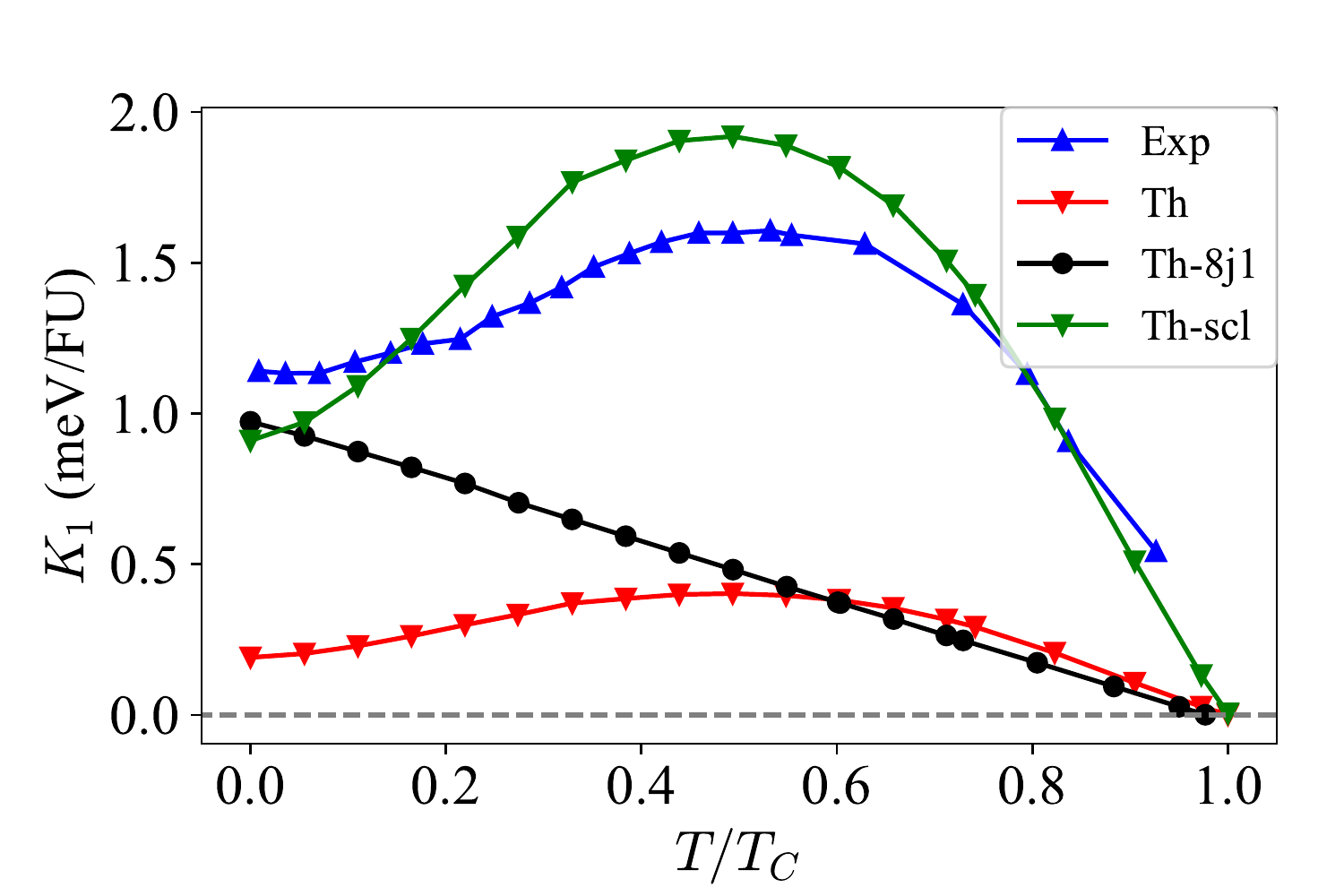}
\caption{The $T$-dependence of the leading anisotropy constant, $K_{1}$, of Y$_2$Fe$_{14}$B from DLM-DFT theory (red curve) and experiment (blue)~\cite{hirosawa1986magnetization}. The green curve shows the theory values, scaled to account for the difference between FP~\cite{jukkr} and ASA (\cite{Note1}) at $T=0$K. The black curve shows $K_{1}$ (unscaled) if the Fe ${8j_{1}}$ sublattice is constrained to be disordered magnetically.}
\label{fig.MAE_temp_YFeB}
\end{figure}

While significantly smaller than from the f-electrons of the Nd atoms, the primarily TM component of the MCA grows in importance with rising $T$. As the RE MCA drops swiftly along with the magnetic order ~\cite{callen1966present,Patrick2020}, the TM contribution can actually increase as shown explicitly in measurements on Y$_2$Fe$_{14}$B~\cite{cadogan1992analysis,hirosawa1986magnetization}. Such non-monotonic temperature variation is puzzling, and has been attributed to a magnetostructural effect from an anisotropic expansion of the crystal lattice, competing single-ion-like contributions~\cite{cadogan1992analysis} or {competing single and two-ion MCA using atomistic spin dynamics simulations~\cite{evans2020temperature,cuadrado2021first}}
Since fully relativistic effects such as spin-orbit coupling are included in our DLM-DFT theory we investigate the MCA temperature dependence directly and show our results in Fig.~\ref{fig.MAE_temp_YFeB} for Y$_2$Fe$_{14}$B.

{Using the highly accurate, full potential (FP) KKR code~\cite{jukkr,Papanikolaou:2002}, we first calculate the MCA at $T=0$K to be $\approx$ 0.9 meV/formula unit (FU) which agrees well with experimental values~\cite{hirosawa1986magnetization}. The same rapid loss of magnetic order with increasing temperature which we find for the Fe 8$j_1$ sites in Nd$_2$Fe$_{14}$B (Fig.~\ref{fig.stuffvT}(b)) is also evident in Y$_2$Fe$_{14}$B~\cite{Note1} and this points to a significant role for this sublattice in the anomalous MCA $T$-dependence. We therefore carry out further FP MCA calculations where now the Fe 8$j_1$ sites are constrained to be magnetically disordered ($m_{8j_1}=0$) via an equal weighting of local moments on each of these sites  along the $\pm x$ and  $\pm z$ directions. The effect on the computed MCA is striking - it increases greatly to $\approx$ 1.7 meV/FU - and we infer that the much faster decrease of 8$j_1$ magnetic order with temperature relative to that on the other Fe sublattices is the key driver for the MCA $T$-dependence.} 

{To test this proposition, we calibrate DLM-DFT MCA values against our $T=0$K FP MCA calculations, given the current implementation with an atomic sphere approximation (ASA). Although the ASA values are smaller than the FP ones, the same large increase of the value when the Fe 8$j_1$ sites are magnetically disordered is found.} In Fig.~\ref{fig.MAE_temp_YFeB} we show the DLM-DFT temperature dependent MCA both using the ASA (red curve) and also scaled by the ratio between the FP and our ASA $T=0$K values (green).
The increase with temperature is evident, peaking at $T/T_c=50\%$ in line with experiment~\cite{hirosawa1986magnetization} {confirming our proposition}. Since the calculations are for a fixed lattice structure, we can exclude thermal expansion as a cause of the non-monotonic behavior. {We also show} the effect on the MCA of forcing the Fe$_{8j_1}$ sublattice to remain magnetically disordered at all $T$, i.e. $m_{8j_1}=0$.
The resulting unscaled MCA, shown in black in Fig.~\ref{fig.MAE_temp_YFeB}, is dramatically altered - the peak has gone and the MCA decays linearly
with temperature and the $T=0$K value is enhanced significantly.
Clearly, establishment of magnetic order on the Fe$8j_1$ sublattice correlates with a substantial drop in the (uniaxial) MCA.

Our ultimate goal is to describe Nd$_2$Fe$_{14}$B's large magnetic anisotropy and its temperature variation. So to the TM MCA we add the dominant RE components. These are calculated~\cite{Patrick_2018,patrick2019first} from the solution of Eq.~\ref{eq.RE} where the crystal field coefficients~\cite{Note1} are  determined from first principles~\cite{Patrick_2019}, and exchange field, $\boldsymbol{B}_\mathrm{exch}$ provided directly by
the DLM-DFT Weiss field for each Nd site (Fig.~\ref{fig.stuffvT}) divided by the computed Nd spin moment of 3.66~$\mu_B$. 

Our calculated exchange fields of 699 and 725~T for the RE $f$ and $g$ sites respectively are somewhat larger than those used in fits of experimental magnetization data (450--520~T~\cite{Loewenhaupt1990}),
but as pointed out in Ref.~\cite{Herbst1991}, the large number of parameters in Eq.~\ref{eq.RE} introduce significant uncertainties.
In principle, INS data would provide a direct measure of the exchange fields but are not available for Nd-Fe-B. Our proposed values are, however, supported by the good
agreement between INS experiments~\cite{Loewenhaupt1991} and our DLM-DFT calculations for the related Gd$_2$Fe$_{14}$B magnet
($324$~T vs $307/319$~T). The Gd exchange fields are substantially smaller than those calculated  for Nd.  The relative difference ($\sim$2) mirrors
that of the spin moments (7.46 vs.\ 3.66 $\mu_B$) and reflects
the similar Weiss fields we calculate for the two materials.
\begin{figure}
\hspace{-5mm}
\includegraphics[width=0.48\textwidth]{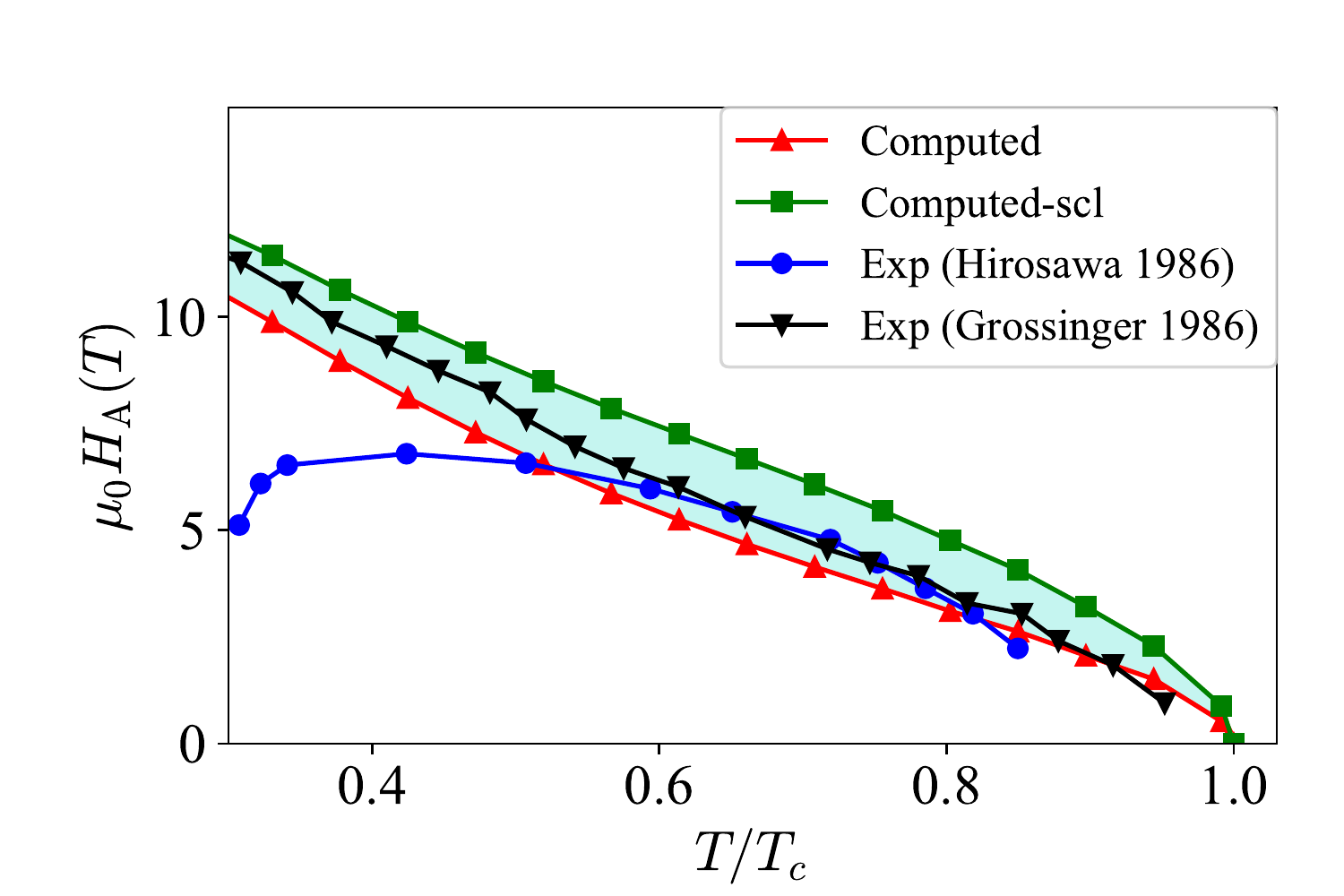}
\caption{{Evolution of the anisotropy field, $H_A$, versus $T/T_c$ from theory compared to experimental 
measurements (from Refs.~\cite{grossinger1986magnetic}, black curve, ~\cite{hirosawa1986magnetization}, blue curve). 
The agreement is good above $T_c/2$. The red and green curves use the non-RE MCA taken from the red and green curves of Fig.~\ref{fig.MAE_temp_YFeB}.}}
\label{fig.MAE_field}
\end{figure}

Using the method of Ref.~\cite{Patrick_2018} we calculate effective
anisotropy constants $K_1(T)$, $K_2(T)$ and anisotropy field, 
$\mu_0 H_A = 2 K_1 / M$ {\it ab initio} which is shown in Figure~\ref{fig.MAE_field}. 
The red/green curves show $\mu_0 H_A$ which includes the non-RE contribution to the MCA of the red/green plots in Fig.~\ref{fig.MAE_temp_YFeB}.  Fig.~\ref{fig.MAE_field} also shows the experimental measurements from
Refs.~\cite{grossinger1986magnetic,hirosawa1986magnetization}.
Below $T/T_c \sim 0.5$ there is some
discrepancy between the two sets of experimental data, but
above there is consistency between both the experiments
and our calculations. The calculations show the clear importance of the Fe-dominated MCA to the anisotropy field at high temperatures - the red curve is over 1~T less than the green one over a range of temperatures despite the contributions from the non-RE MCAs differing by less than 30~$\mu$eV per Fe atom. 

Nd$_2$Fe$_{14}$B's spin reorientation transition (SRT) at 135K~\cite{Yamada1988,Herbst1991,Toga2016} is not described by our calculations owing to an underestimate of the high order crystal field coefficients~\cite{Hummler1996,Toga2016,Tatetsu2018}. This shortcoming exemplifies a more general challenge for theory modelling of low $T$ strongly correlated f-electron effects to construct a robust way to significantly enhance the value of these coefficients~\cite{Pourovskii2020}. Around room temperature and above, however, the effects on the MCA from these high order terms are small. This is also the temperature regime where the tenets of our DLM-DFT theory are valid. 

Nd$_2$Fe$_{14}$B and the RE-TM permanent magnet family to which it belongs have a compelling set of attributes. Their technological value is enormous and growing and their magnetic properties, at a fundamental level, come from a rich and subtle combination of RE, localized, and TM, itinerant electron, effects. To enhance magnetic functionality and extract pointers for the development of even better materials, multiple interrelated aspects have to be accounted for.  Our {\it ab initio} DLM-DFT modelling has shown the importance of describing accurately the rich and complex itinerant electron magnetism associated with the Fe sites and valence electrons generally for the production of the robust exchange field acting on the RE atoms, the higher order effective spin interactions and the nature of the non-f electron MCA. 
The modifications proposed here should be incorporated into future
atomistic, effective spin and micromagnetic modelling to correctly
describe these phenomena.

The work was supported by  EPSRC (UK) Grant No. EP/M028941/1 (J.B. and J.B.S.) and  Royal Society Research Grant RGS\textbackslash R1\textbackslash201151 (C.E.P.).
\bibliography{papers}

\end{document}